# Nature of the bound states of molecular hydrogen in carbon nanohorns

**Félix Fernandez-Alonso,**
*ISIS Facility, Rutherford Appleton Laboratory, Chilton, Didcot, Oxfordshire OX11 0QX, United Kingdom*

**F. Javier Bermejo, Carlos Cabrillo,**
*Instituto de Estructura de la Materia, C.S.I.C., and Dept. Electricidad y Electrónica-Unidad Asociada CSIC, Facultad de Ciencia y Tecnología, Universidad del País Vasco / EHU, P.O. Box 644, E-48080-Bilbao, Spain*

**Raouf O. Loutfy,**
*MER Corporation, 7960 South Kolb Road, Tucson, Arizona 85706, USA*

**Vincent Leon, Marie-Louise Saboungi**
*Centre de Recherche sur la Matière Divisée, C.N.R.S - Université d'Orléans, 1B, rue de la Férollerie, 45071 Orléans Cedex 2, France*

## ABSTRACT

The effects of confining molecular hydrogen within carbon nanohorns are studied via high-resolution quasielastic and inelastic neutron spectroscopies. Both sets of data are remarkably different from those obtained in bulk samples in the liquid and crystalline states. At temperatures where bulk hydrogen is liquid, the spectra of the confined sample show an elastic component indicating a significant proportion of immobile molecules as well as distinctly narrower quasielastic line widths and a strong distortion of the line shape of the *para→ortho* rotational transition. The results show that hydrogen interacts far more strongly with such carbonous structures than it does to carbon nanotubes, suggesting that nanohorns and related nanostructures may offer significantly better prospects as light-weight media for hydrogen storage applications.

# I. INTRODUCTION

After a rather optimistic start [1], the usefulness of carbon-based nanostructured materials (mostly in the form of multi-, MWNT, or single-walled, SWNT, nanotubes) as viable hydrogen storage media remains questionable. While early work [1] reported values largely in excess of the US-DOE target of 6.5 %, most subsequent studies could not verify the promising prospects referred to in Ref. [1]. The difficulty in reproducing such measurements has usually been attributed to problems in preparing SWNT or MWNT samples of high enough purity [2]. One study [3] has shown that molecular hydrogen ($H_2$) loaded into SWNTs at relatively high pressures (~100 bar) behaves as free $H_2$, as attested by the absence of any significant change in the line profile of the *para→ortho* rotational transition, while another investigation [4] reports a rather different behavior. Moreover, $H_2$ loaded into carbon nanotubes at low temperatures does not bind firmly to the carbonous matrix and most of it evaporates once the sample is brought back to room temperature. This latter fact severely limits the potential applications of carbon nanotubes since any useful material must be able to release adsorbed $H_2$ both under controlled conditions and mild thermal treatments.

Single-walled carbon nanohorns (SWNH) consist of graphitic structures formed out of a single-walled graphene sheet with an average size of 2-3 nm. They adopt a horn-like shape and aggregate to form flower-like structures with sizes of about 80-100 nm. These nanostructures exhibit very large surface areas approaching 1500 $m^2g^{-1}$ [5] and are therefore attractive candidates for gas and liquid storage. Moreover, no metal catalyst is required for their synthesis thus enabling low-cost, large-scale production of high-purity samples.

Our interest in these materials stems from previous studies [8] where $H_2$ adsorption isotherms suggested a higher adsorbate density than that of the liquid at 20 K, with values approaching those of solid $H_2$ near its triple point (T = 13.83 K). This remarkable solid-like behavior was attributed to strong quantum effects [8], a conjecture later confirmed by detailed quantum simulations [9]. More recently, Tanaka *et al.* [10] have reported isosteric heats of adsorption for $H_2$ on SWNHs nearly three times as large as those on SWNTs, corresponding to $H_2$ binding energies as high as 100-120 meV. This increase in $H_2$ binding energy was attributed to strong solid-fluid interactions at the conical tips. Weaker $H_2$ adsorption was also observed away from the SWNH tips, with an interaction very similar to what has been found for SWNTs. In addition, preliminary results on $H_2$ adsorption in SWNHs [6] have shown that the *para→ortho* rotational line shape is significantly distorted from what is observed in SWNT and bulk $H_2$ samples. Also, thermal cycling from 10 K to 290 K and back to 10 K recovers a spectrum close to that measured after the initial low-temperature loading, indicating that most $H_2$ remains
firmly attached to these carbon nanostructures. In the present Letter, high-resolution neutron spectroscopy is utilized to gain unique microscopic insight on the nature of $H_2$ adsorption on SWNHs.

## II. RESULTS AND DISCUSSION

Quasielastic (QENS) and inelastic (INS) neutron scattering are powerful probes of the dynamics and bonding of molecules in condensed matter [11]. Here we report on QENS and INS experiments of $H_2$ physisorbed on SWNHs. Our aim is to probe the bound states of $H_2$ in these novel carbon nanostructures via an analysis of the dependence of the QENS line widths and intensities on wave vector and temperature. Deviations of the *para→ortho* INS rotational line shape from free-rotor behavior provide additional insight on the strength and angular character of the interactions between $H_2$ and the SWNH surface.

SWNH samples were obtained from MER Corporation (Tucson, Arizona, USA). The IRIS backscattering spectrometer at ISIS represents an optimal choice for these experiments as it can provide high-resolution QENS and INS data [12]. The instrument configuration was set to the 002 reflection from cooled pyrolytic graphite with a beryllium filter to prevent contamination from higher-order reflections. QENS measurements covering an energy transfer range of -0.2 − 1.0 meV with an energy resolution of 8.8 $\mu$eV (HWHM) were carried out at T = 5, 15, and 25 K. INS spectra were recorded at T = 1.5 K over the energy-transfer range 6 - 20 meV by using the same energy analyser and phasing the choppers to run in offset mode, providing an energy resolution of 75 $\mu$eV at energy transfers around the $H_2$ *para→ortho* rotational transition at 14.7 meV. For the adsorption experiments, *normal* $H_2$ (75% ortho, 25% para) was condensed at T = 25 K and P = 1 bar into an annular-geometry aluminum cell. In order to keep total scattering powers below the canonical value of 15 %, sample thicknesses of 0.25 mm and 1 mm were used for bulk and adsorbed $H_2$, respectively.

Fig. 1 displays a comparison of QENS spectra for the bulk liquid (L) as well as $H_2$ adsorbed on SWNHs ($H_2$NH). An obvious difference between the two sets of data is the strictly elastic component in $H_2$NH, indicating the presence of species with highly reduced mobility at temperatures and pressures where bulk $H_2$ is liquid.

QENS spectra were first analysed with a model-free Bayesian algorithm that infers from the data the mini- mum number of spectral components needed to account for the experimental observations [13]. The results show that the spectra for L can be described adequately with a single Lorentzian component, whereas those for $H_2$NH require an additional, resolution-broadened elastic line. The wave vector dependence and the intensity of the quasielastic components for both L and $H_2$NH are displayed in Fig. 2.

QENS data were analysed further by recourse to a jump-diffusion model where the HWHM of the QENS width is given by

$$\Gamma(Q) = \frac{D_T Q^2}{1 + \frac{D_T Q^2}{E_0}} \qquad (1)$$

where $D_T$ stands for the translational self-diffusion coefficient and $\tau = \hbar/E_0$ is a mean residence time.

Our value for the self-diffusion coefficient of L, $D_T$ = 0.43±0.01 Å$^2$ ps$^{-1}$, compares favourably with 0.47 Å$^2$ ps$^{-1}$ of Egelstaff *et al.* [14] and with a value of 0.43±0.09 Å$^2$ ps$^{-1}$ obtained by constant-gradient spin-echo NMR [15]. As shown in Fig 2, QENS intensities in L conform to the form factor of a freely rotating $H_2$ molecule given by [16]

$$I_{rot}(Q) = j_0^2(QR_e/2) + 2j_2^2(QR_e/2) \qquad (2)$$

where $j_n$ stands for spherical Bessel functions and $R_e$ = 0.7416 Å is the internuclear separation in $H_2$ [19]. The analysis of the QENS spectra for $H_2NH$ required an additional delta-function contribution with an intensity about four times that of the lorentzian component. The elastic line proves that a large fraction of molecules are immobile within the time window of the measurements (1 – 75 ps), while the QENS line can be assigned to translational motions of the mobile fraction. The wave-vector dependence of the latter at T = 15 K and T = 25 K is shown in Fig 2. The higher temperature is about 5 K above the boiling point of the liquid at a pressure of 1 bar and so what we are observing here corresponds to a fluid tightly bound to the SWNH matrix. Analysis of the QENS energy widths yields values for $D_T$ of 0.96±0.1 Å$^2$ ps$^{-1}$ and 6.5±1.7 Å$^2$ ps$^{-1}$ at T = 15 K and 25 K, respetively. Values for the characteristic energies $E_0$ of 0.13±0.01 meV and 0.55±0.045 meV at 15 K and 25 K, respectively, are to be compared with 3.38±0.27 meV for L at T = 15 K. These results show that the mobile fraction of $H_2$ undergoes jump motions that are some 2.3 times faster than those found in L at the same temperature, whereas the residence time increases by a factor of about 25 due to interaction with the carbon matrix. Corresponding estimates for the length of a diffusive step $l = \sqrt{6D\tau}$ are 0.7±0.07 Å and 5.44±0.41 Å for L and $H_2NH$ at T =15 K, respectively, and 6.81±0.56 Å for $H_2NH$ at T = 25 K.

Further information about the bound $H_2$ fraction is provided by the elastic form factors. Results are presented in Fig 3. Such data may be parameterized using an equation similar to Eq. 2 with the inclusion of an extra multiplicative factor to read $I_{elast}(Q) = I_{rot}(Q) \cdot I_{cm}(Q)$. The Debye-Waller factor $I_{cm} = \exp(-<u^2>Q^2/3)$ is characterized by an effective mean-square amplitude $<u^2>$ describing motions of the $H_2$ center of mass. In the present description, it is the only free parameter, since the internuclear distance $R_e$ = 0.7416 Å has been fixed to its tabulated value. Least-squares fitting of the elastic intensity data to the above expression reveals an increase of the effective $<u^2>$ from 0.77±0.06 Å$^2$ at T = 5 K to 1.47±0.08 Å$^2$ at T = 25 K. The observed temperature dependence of $<u^2>$ may be further understood in terms of a harmonic oscillator with energy levels $E_n = \hbar\omega_c(n + 1/2)$ and Bose population factors $P_n(T) = \exp(-\beta E_n)/\sum_n \exp(-\beta E_n)$ according to [17]

$$I_{elast}(Q) = I_{rot}(Q) \cdot \sum_n P_n(T) e^{-<u_n^2>Q^2/3} \qquad (3)$$

where $<u_n^2> = (n + 1/2)\hbar/M\omega_c$, M is the mass of $H_2$, $\omega_c$ is a characteristic frequency of centre-of-mass vibration, and $\beta = 1/k_BT$. From our lowest-temperature measure-ments at T = 5 K we obtain $\omega_c$ = 1.36 meV. As shown in Fig. 3, calculation of the expected form factors at higher temperatures using this value of $\omega_c$ provides a unified description of the data.

The H$_2$NH INS spectrum displays a severely distorted line shape suggestive of inhomogeneous broadening, as shown in Fig. 4. Moreover, the broad component is markedly asymmetric with a bias towards lower frequencies. This situation represents a stark departure from SWNTs, where H$_2$ spectra at T = 4.5 K display a single and narrow contribution centered at 14.5±0.1 meV indicative of free rotation [3]. These differences are also in agreement with the higher binding energies found by Tanaka *et al.* [10]. To analyze our results, we consider a Stark splitting of the *para→ortho* rotational transition. Physically, it could arise from the presence of electrostatic interactions between H$_2$ and the carbon matrix. To lowest order, this interaction potential takes the form $V(\Theta, \Phi) = V_2 \cos^2\Theta$, where $V_2$ stands for the height of the potential barrier and $\Theta$ is the angle between the diatom axis and the field direction (e.g., the surface normal) [18]. Using this model, the *para→ortho* transition energies have been calculated as a function of barrier height by direct diagonalization of the Hamiltonian matrix in the free-rotor $|JM> \equiv Y_{JM}(\Theta, \Phi)$ basis, where $Y_{JM}(\Theta, \Phi)$ is a spherical harmonic, and a given $|JM>$ state is characterized by a rotational energy $E_{JM} = B_{rot}J(J + 1)$. Convergence of energy eigenvalues was ensured by using a basis-set size $J \geq 20$. The results of these model calculations are shown in the inset of Fig. 4. The gas-phase value $B_{rot}$= 7.35 meV [19] has been assumed throughout the analysis. The three Zeeman sublevels of the ortho state split into two components corresponding to the M = ±1 (maximal) and 0 (minimal) projections of the total angular momentum along the quantization axis. In the limit of a small perturbation (e.g., $V_2/B_{rot} \to 0$), both intensities $I_M$ and energy positions relative to the free-rotor energy $\Delta E_M$ approach a constant ratio given by $I_{M=\pm 1}/I_{M=0} = \Delta E_{M=0}/\Delta E_{M=\pm 1} = 2$. By constraining our fitting procedures to obey these restrictions, it is possible to fit the observed INS spectrum as shown in Fig. 4. The unperturbed rotational transition gives rise to the narrow component centered at 14.68±0.01 meV with a resolution-limited HWHM of 0.07±0.01 meV. In addition, two broad satellites are centered at 14.51±0.03 meV and 15.02±0.04 meV with HWHMs of about 0.31 meV. The intensity ratio of the Stark-split component relative to the unperturbed line is roughly 3.3, that is, about 75 % of the molecules are interacting with the carbon lattice strongly enough to perturb their high-frequency internal rotation. This ratio agrees with the amount of immobile H$_2$ extracted from QENS data, further reinforcing the notion that the most energetically favourable SWNH adsorption sites lead to a solid-like H$_2$ phase characterized by a significant angular anisotropy in the interaction potential. The observed INS spectral splitting is about 0.5 meV, corresponding to a temperature of about 6 K and a hindering barrier in the meV range. Moreover, since most of the spectral intensity appears below the free-rotor line, the preferred orientation of the H$_2$ molecular axis must be parallel to the SWNH surface.

Lacking detailed information on the phonon spectrum of SWNH, an assignment of the vibrational frequency $\hbar\omega_c$ ~ 1.3 meV must remain tentative. This frequency is certainly below the phonon spectrum of molecular hydrogen [20] so it must be related to low-energy modes of the SWNH substrate. In spite of its relevance to explain our experimental findings quantitatively, we note that the role of carbon-substrate motions has been largely neglected in simulation work to date [7, 9, 10]. Also, since $\hbar\omega_c$ ~ $V_2$, the coupling between internal (rotational) and center-of-mass (translational) degrees of freedom is likely to play an important role in dictating the precise nature of $H_2$ motions on the SWNH substrate. In this context, recent $^{13}C$ NMR relaxation data [21] have identified two distinct processes with disparate relaxation times. On the basis of previous information on the electronic subsystem [22], these can be assigned to either fast processes occurring in SWNHs in close proximity to the surface of the bundles or to slow motions taking place inside the graphitic cores.

## III. CONCLUSION

In summary, this Letter provides unambiguous experimental signatures of a strong interaction between $H_2$ and SWNHs. More importantly, the character of this interaction is quantitatively different from what is known for $H_2$ in SWNTs. Both the stochastic (i.e., mass diffusion) and vibrational dynamics of free and bound species are strongly altered due to their adsorption on the carbon substrate. The mobile fraction exhibits a remarkably higher mobility than in the weakly interacting bulk phase. The vibrations of the SWNH−$H_2$ complex also display large mean-square displacements, indicative of strong quantum effects that persist at temperatures well above the boiling point of the bulk liquid. Our results also explain findings from previous modeling and experimental studies reporting the presence of two kinds of preferential adsorption sites with significantly different mobilities [10].

## AKNOWLEGMENTS

We thank RBE Down and CM Goodway from the ISIS User Support Group for their expert assistance. FFA gratefully acknowledges financial support from the UK Science and Technology Facilities Council.

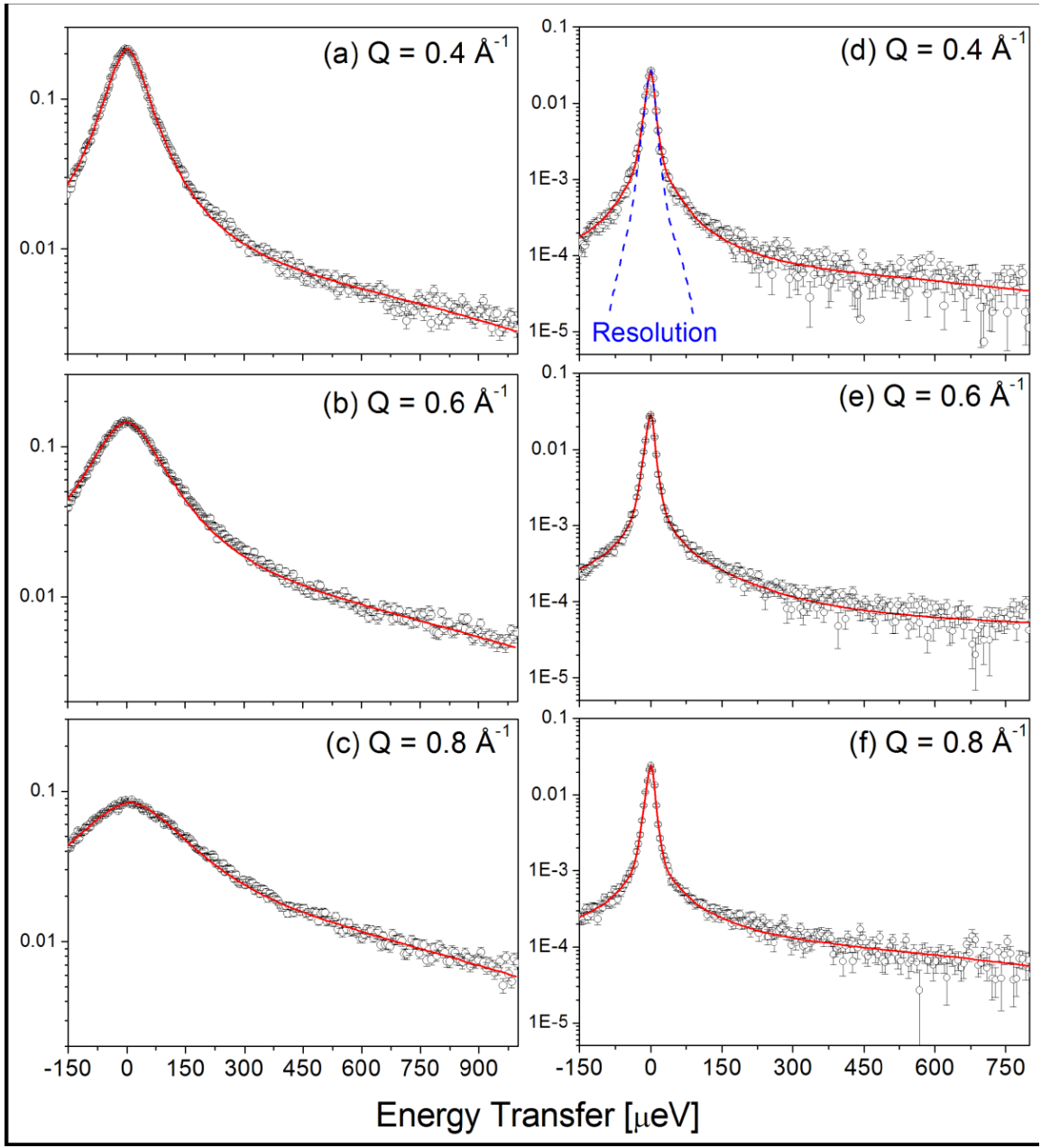

FIG. 1: QENS spectra at various wave vectors Q for the L (a-c) and H$_2$NH samples (d-f) at T = 15 K. The solid lines are model fits to the data. The narrow line in (d) represents the instrumental energy resolution.

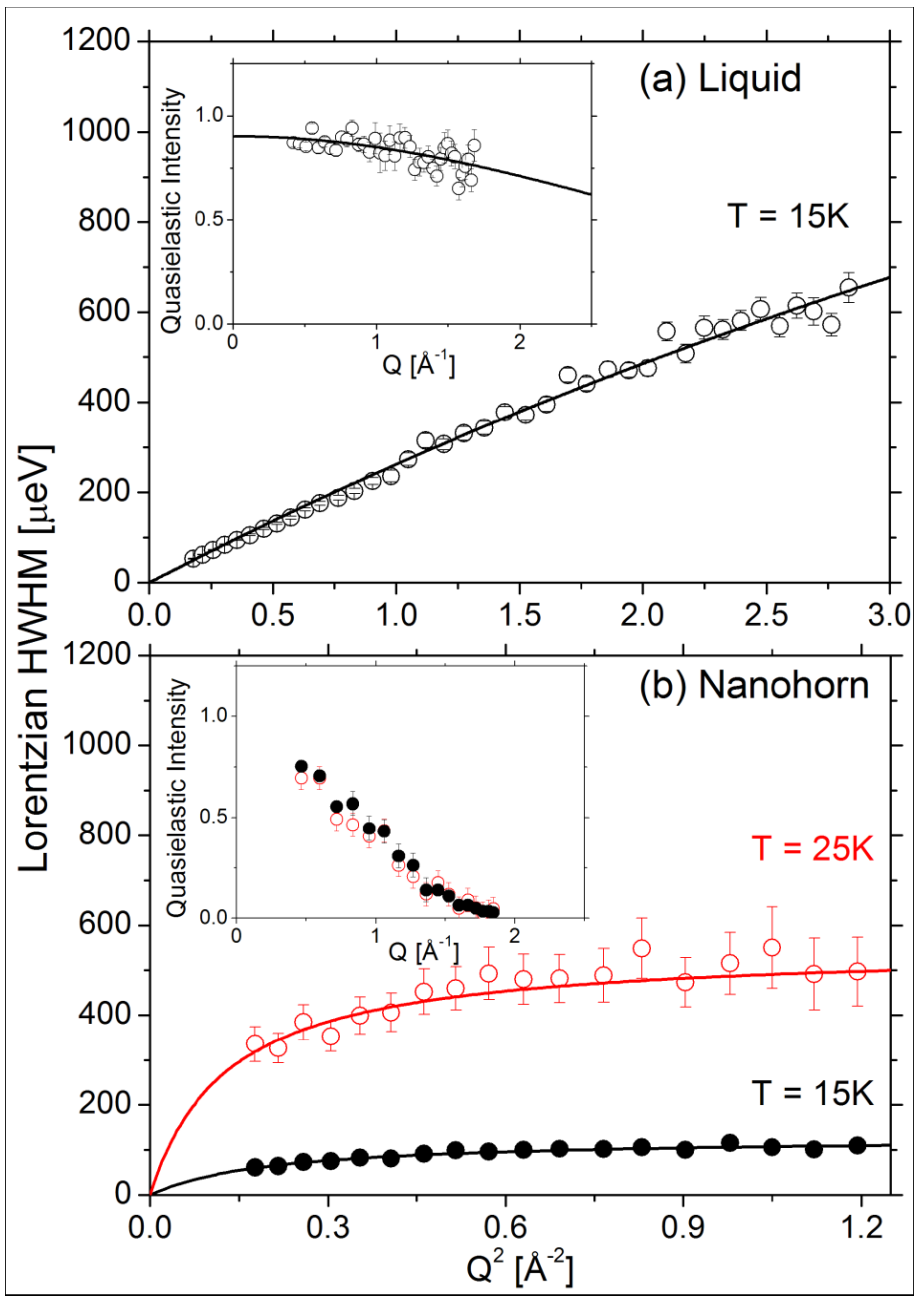

FIG. 2: Wave-vector dependence of the QENS widths and amplitudes (insets) for L (top) and $H_2NH$. Solid lines are fits to these data as explained in the text.

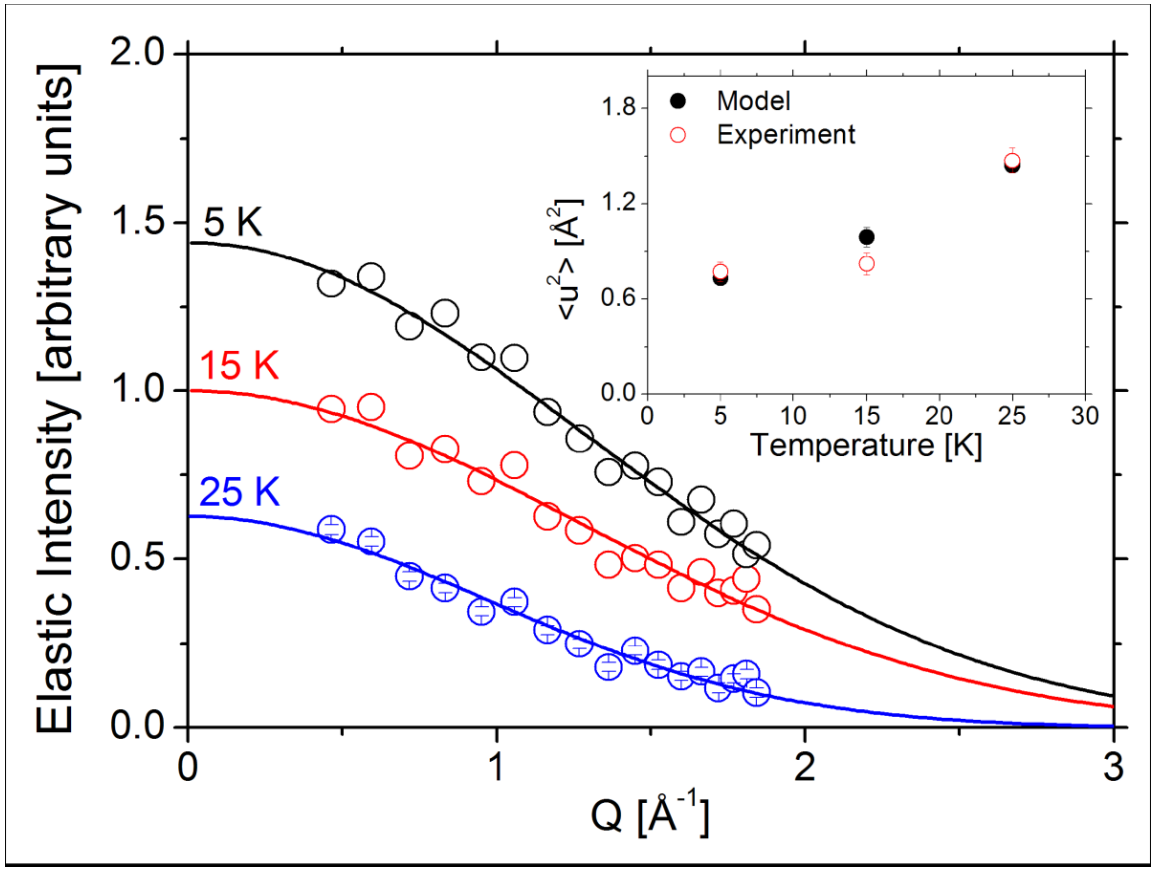

FIG. 3: Wave-vector dependence of the elastic intensity in $H_2NH$. Solid lines are fits using Eq. 3. The inset shows the temperature dependence of the expected and observed values for the effective $<u^2>$.

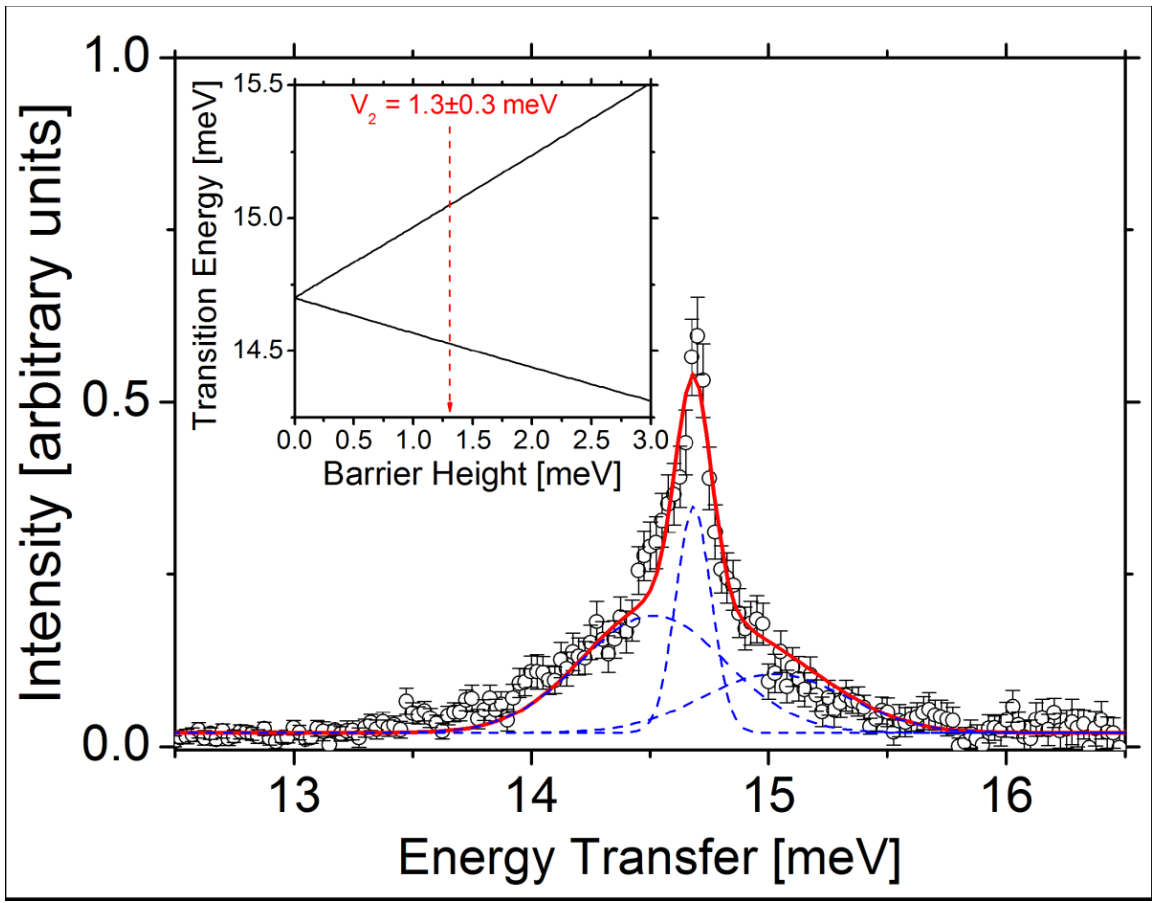

FIG. 4: H$_2$NH INS spectra at T = 1.5 K. The inset shows a schematic diagram of the M-dependent energy-level splittings described in the text.